\documentclass[%
 reprint,
 amsmath,amssymb,
 aps,
]{revtex4-2}

\usepackage{graphicx}
\usepackage{color}
\usepackage{dcolumn}
\usepackage{bm}
\usepackage{amsmath}
\usepackage{amssymb}
\usepackage{mathrsfs}

\begin{document}

\preprint{APS/123-QED}

\title{
Reinforcement Learning Approach to Shortcuts between Thermodynamic States
with Extra Constraints}

\author{Rongxing Xu}
\email{xurongxing@keio.jp}
\affiliation{%
 Department of Physics, Keio University, 3-14-1 Hiyoshi, Yokohama 223-8522, Japan
}
\affiliation{%
 Mathematical Science Team, RIKEN Center for Advanced Intelligence Project (AIP), 1-4-1 Nihonbashi, Chuo-Ku, Tokyo 103-0027, Japan
}%


\date{\today}

\begin{abstract}
We propose a systematic method based on reinforcement learning (RL) techniques to find the optimal path that can minimize the total entropy production between two equilibrium states of open systems at the same temperature in a given fixed time period. Benefited from the generalization of the deep RL techniques, our method can provide a powerful tool to address this problem in quantum systems even with two-dimensional continuous controllable parameters. We successfully apply our method on the classical and quantum two-level systems.
\end{abstract}

\maketitle


\section{\label{sec:level1}Introduction}
Problems on accelerating transitions between two states commonly appear in many physical situations. Understanding the transition mechanism between two states and developing accelerating methods are prerequisites to upgrade modern technology and physics. In the quantum mechanical time-evolution, the topic of shortcut to adiabaticity has been intensively studied to find protocols for transitions between ground states \cite{berry2009transitionless, torrontegui2013shortcuts,guery2019shortcuts,del2019focus}. The technique and concept have been successfully applied and extended in providing better protocols in quantum computing \cite{hegade2021shortcuts,santos2015superadiabatic,takahashi2019hamiltonian}, optimizing the atom cooling \cite{torrontegui2011fast,chen2010fast,du2016experimental} and improving the performance of microscopic heat engines \cite{hartmann2020many,deng2013boosting,beau2016scaling,tobalina2019vanishing}. Similarly, finding shortcuts in thermodynamic transformations is also a fundamental problem. The thermodynamic transformation here includes the thermalization to obtain an equilibrium distribution from an initial state \cite{jing2013inverse,mukherjee2013speeding,dann2019shortcut,dann2020fast}, and the acceleration of isothermal processes \cite{li2017shortcuts,albay2019thermodynamic,albay2020realization}, and so on. In the thermalization problem, Dann et al. proposed a method to find a shortcut of the transitions between equilibrium states \cite{dann2019shortcut} in the quantum case. Mart{\'i}nez et al provide a so-called engineered swift equilibration protocol to deal with such problem in the classical cases \cite{martinez2016engineered,le2016fast,chupeau2018engineered}. Another intriguing direction is the shortcut to isothermality (ScI), where they accelerates the isothermal process under the condition that the system always stays in the instantanuous equilibirum states \cite{li2017shortcuts,albay2019thermodynamic,albay2020realization}.  

\begin{figure}[t]
\includegraphics[width=1\columnwidth]{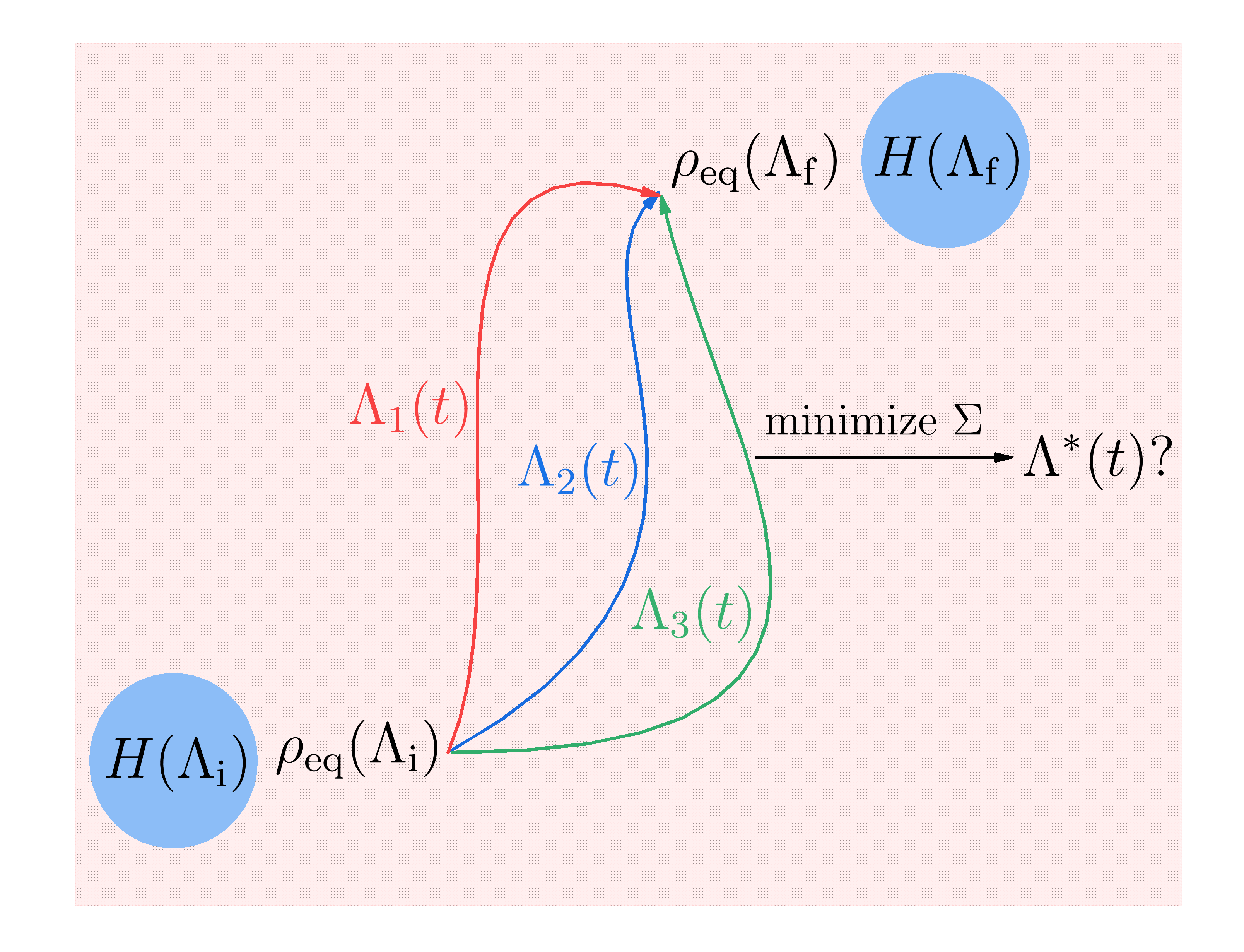}
\caption{\label{fig:0} The schematic graph of the shortcut between two equilibrium states. The system (blue circle) evolves in a thermal environment (pink background) with a constant temperature $T$ in a given fixed time period. An optimal protocol $\Lambda^*(t)$, which minimizes the entropy production $\Sigma$ during the transition between two equilibrium states $\rho_{\rm eq,i}$ and $\rho_{\rm eq,f}$, is required to be obtained.}
\end{figure}

Here, we note that in any acceleration protocol, extra time-dependent perturbation is necessary. This implies that additional energy costs have to be accompanied. This topic has been noticed recently when designing the ScI protocols in both quantum \cite{pancotti2020speed} and classical systems \cite{albay2019thermodynamic}. As finite-time thermodynamic process can be applied to improve the performance of thermodynamic cycles of microscopic heat engines, understanding the associated dissipative energy is of great practical importance. From these backgrounds, it is necessary to develop a systematic method to find an optimal protocol that satisfies i) accelerating the transitions between two states, and ii) minimizing additional dissipation arising from the accelerated transition. As it is very difficult to incorporate the methods so far, a general approach is desired. Reinforcement learning (RL) is considered to be a promising framework to deal with our problem. By regarding the optimization process as a series of movements of an intelligent agent in an environment in order to maximize the cumulative reward \cite{sutton2018reinforcement}, it has been proved to have successful applications in quantum protocol design \cite{paparelle2020digitally,porotti2019coherent,bukov2018reinforcement}, ground state searching in quantum many-body systems \cite{carleo2017solving,deng2017machine} and finding optimal protocols that minimize the entropy production of open quantum systems \cite{sgroi2021reinforcement}.

In this paper, we step forward in this direction using RL method for an acceleration of thermodynamic transformations. To this end, we consider a simple problem finding a shortcut between two equilibrium states. We aim to obtain the equilibrium distribution of some system parameter $\Lambda_{\rm f}$ when the initial state is an equilibrium state with a parameter $\Lambda_{\rm i}$ with the fixed time period $\tau$ in open classical as well as quantum dynamics (See Fig. \ref{fig:0} for the scheme). Since it is in general difficult to control thermal reservoirs, we fix the dissipative dynamics. We control only system parameters of the path: $\Lambda_{\rm i} \to \Lambda_{\rm f}$, so that dissipation becomes minimum within the whole paths making the final state in equilibrium with a parameter $\Lambda_{\rm f}$.
In this paper, We propose a method of the deep reinforcement learning techniques, specifically, the policy gradient technique combined with deep neural networks. 
We show that our method works well, providing a powerful tool to address this problem in quantum systems, even when the controllable parameters are two-dimensional, which has not been reported before to the best of our knowledge. 

This paper is organized as follows. We first formally explain our main purpose in Section \ref{sec:level2} and briefly review the RL method and apply it to our problem in Section \ref{sec:level3}. We then explain the dynamics of the two-level system and give a theoretical analysis of our problem in classical case in Section \ref{sec:level4}. The results of our RL method is disscussed in Section \ref{sec:level5}.

\section{\label{sec:level2}Main purpose}
We first explain our main purpose. We consider the open system which includes a system with Hamiltonian $H(\Lambda)$ and a thermal environment with a constant temperature $T$, as shown in Fig. \ref{fig:0}. Here, $\Lambda$ is a set of time-dependent controllable parameters and it can vary continuously. 

Now we consider the following transition process. First, we assume that the system initially stays at some instantaneous equilibrium state $\rho_{\rm eq,i}$, 
\begin{equation}
    \rho_{\rm eq,i} = \frac{e^{ - \beta H (\Lambda_{\rm i}) }} {Z_{\rm i}} \, ,  ~~~
Z_{\rm i} =   
    {\mathrm{Tr} \left[e^{-\beta H (\Lambda_{\rm i})} \right]} \, , \label{eq1}
\end{equation}
where we denote $\beta$ to be the inverse temperature and set the Boltzmann constant to be unity. Then, by keeping the temperature $T$ constant, we operate the system by controlling $\Lambda$ from the initial equilibrium state $\rho_{\rm eq, i}$ at the time $t_{\rm i}$ to another equilibrium state $\rho_{\rm eq,f}$ at the time $t_{\rm f}$. We aim to search for an optimal path $\Lambda^*(t)$ that can minimize the entropy production $\Sigma$ during the whole transition process above (See Fig. \ref{fig:0} for the schematic graph), i.e., 
\begin{equation}
\begin{aligned}
    &\text{arg} \min_{\Lambda(t)}  \Sigma = \int_{t_{\rm i}}^{t_{\rm f}} {\rm d} t\, \dot{\Sigma}  \\
   & ~~\text{s.t.}~~  \Delta d = d(\rho(t_{\rm f}), \rho_{\rm eq,f}) = 0 \, , \\ 
    &~~~~~~~~~~\,\tau=t_{\rm f} - t_{\rm i}: {\rm fixed}\label{eq2}
\end{aligned}
\end{equation}
where $\rho(t_{\rm f})$ is the final state at time $t_{\rm f}$ and $d(\rho_1, \rho_2)$ measures the distance between to different states, where we use trace distance (${\rm L}_1$ norm) for quantum (classical) systems.
In general, it is difficult to solve this type of optimization problems exactly, we here employ the reinforcement learning (RL) techniques in subsequent sections.

\section{\label{sec:level3}Method}
In this section, we explain our RL method. 
Generally speaking, RL is one of the machine learning technique where a computer agent learns to accomplish a designated task by executing a series of actions aiming at maximizing the total reward induced by the interaction with the environment \cite{sutton2018reinforcement}. The process can be divided into a series of discrete steps. At the $j$-th step, the agent observes the environment and obtain its current state $s_j$. Then, according to $s_j$, the agent determines which action $a_j$ is better to maximize the total reward. After executing $a_j$, the agent interacting with the environment makes the current state transform to another state $s_{j+1}$ and receives a reward $r_j$. Repeating the step until it arrives the terminal time or state. We call this whole process is an episode. After several episodes, our agent can learn how to maximize the total reward $R=\sum_j r_j$ with the RL algorithms. 

According to the RL theory, the whole state-action-reward episode can be regarded as a Markovian decision process (MDP). All states, actions, rewards and transition probabilities at every time step in an episode respectively construct the state set $\mathcal{S}$, the action set $\mathcal{A}$, the reward set $\mathcal{R}$ and transition probability set $\mathcal{P}$ of the MDP. To realize an RL technique, it is crucial to determine $\mathcal{S}$, $\mathcal{A}$, $\mathcal{R}$ and $\mathcal{P}$.

For our optimization problem (\ref{eq2}), because the next state is determined once the current state and the action has been determined, the transition is always deterministic. Now, we divide the duration time $\tau$ ($=t_{\rm f}-t_{\rm i}$) into $N$ intervals and assume that there is a computer agent which can obtain the state $\rho(t_j)$ of the system at the $j$-th step of an episode. The state set is constructed by all possible states of the system,
\begin{equation}
    \mathcal{S} = \{s_j = \rho(t_j)\, |\text{~for~all~} j \in [1,N] \} \, . \label{eq3}
\end{equation}
Similarly, the action set is defined as the collection of all possible controllable parameter vectors $\Lambda(t_j)$, i.e.,
\begin{equation}
    \mathcal{A} = \{a_j = \Lambda(t_j)\, |\text{~for~all~} j \in [1,N] \} \, . \label{eq4}
\end{equation}
Most importantly, choosing an suitable quantity as the reward can determine direction of evolution of the MDP and finally affects the performance of the RL. Generally speaking, for an optimization problem, the reward for each step is considered according to the contribution it can provide to the target function. In our problem, noticing that we have two major goals, minimization of the entropy production and arriving at the final state, needs to be simultaneously accomplished, the reward is designed as a linear combination of these two goals,
\begin{equation}
\begin{aligned}
    &\mathcal{R} = \{r_j = (1-\zeta) r_{1,j} + \zeta r_{2,j} | \text{~for~all~} j \in [1,N]\} \, , \\
    &\text{where}~~ r_{1,j} = -\dot{\Sigma_j},~~
    \text{and}~~ 
    r_{2,j} = - \delta_{N,j} \Delta d \, . \label{eq5}
\end{aligned}
\end{equation}
$\delta_{N,j}$ is the Kronecker delta function and $\zeta$ is a free constant varying from different problems. We choose $\zeta$ to be $0.9$ in our following discussions as the optimal parameter here. Here $\Delta d$ is defined in Eq.(\ref{eq2}).

Then, the action at each step is given by the so-called policy function $\pi(a|s)$ which determines the probability to executing the action $a$ when the current state is $s$. The aim of RL is to find the optimal $\pi(a|s)$ to maximize the total reward. According to the theory of policy gradient \cite{sutton2018reinforcement}, we parametrize policy function $\pi(a | s)$ with the parameter vector $\bm{\theta}$, i.e., $\pi(a | s, \bm{\theta})$. Without loss of generality, in this paper, we choose the following parametrized multivariate normal distribution to be the policy function:
\begin{equation}
\begin{aligned}
    \pi(a | s, \bm{\theta}) = &
    \frac{1}{ (2\pi)^{d/2} |\Sigma_a|^{1/2}} \\ \times & \exp{\left[
    -\frac{1}{2}
    (a - \mu_{\theta}(s))^T
    \Sigma_a^{-1}
    (a - \mu_{\theta}(s))
    \right]} \, , \label{eq10}
\end{aligned}
\end{equation}
where $\bm{\theta}$ is the parameter vector and $\Sigma_a$ is the covariance matrix of the action $a$. Here, the parametrized expectation value of $a$ over the distribution ($\mu_{\theta}(s)$) is represented by an neural network for general cases. According to the dimension $d$ of action $a$, Eq. (\ref{eq10}) has different forms, which is specified by the model. Then, the aim of the RL is transformed into finding the optimal $\bm{\theta}$ that can maximize the total reward.

The policy gradient theorem indicates the update rule \cite{sutton2018reinforcement}:
\begin{equation}
    \Delta \bm{\theta} = \alpha G_j \nabla \ln \pi(a_j | s_j, \bm{\theta}) \, , \label{eq9}
\end{equation}
where the notation $G_j$ is the expected return of the $j$-th step: 
\begin{equation}
    G_j = \sum_{k=j}^N r_{k} ~~\text{for}~~ j \in [1,N] \, . \label{eq6}
\end{equation}

Combining (\ref{eq10}), (\ref{eq9}) and (\ref{eq6}), the update rule of the parameter $\bm{\theta}$ of this neural network can be then derived. After enough number of episodes, $\bm{\theta}$ will be optimized and the optimal actions can be taken according to (\ref{eq10}). Please see Appendix A for technique details and hyperparameters.
We emphasize that using multivariate normal  distribution (\ref{eq10}) is important in our method, because it enable us to control multiple parameters simultaneously.

\section{\label{sec:level4}The two-level system}
The example model for performing our main purpose with RL is the two-level system attached to the environment with a constant temperature $T$. We consider the following Hamiltonian for the two-level system:
\begin{equation}
    H(\Lambda_t) = \frac{1}{2} \left( \varepsilon_t \sigma_x + \lambda_t \sigma_z \right) \, , \label{eq11}
\end{equation}
where $\sigma_x$ and $\sigma_z$ are respectively the $x$ and $z$ component of the Pauli matrices. The $x$-component of Zeeman term gives the quantum coherence effect for the eigenbasis for the $z$-component of the Zeeman term. Both $\varepsilon$ and $\lambda$ changes continuously in time. We denote the parameters  $\Lambda_t = (\varepsilon_t, \lambda_t)$ with the vector notation for convenience. Besides, we set $\hbar$ to be unity.

For the dynamics of the system attached to thermal environment, we use the Gorini-Kossakowski-Lindblad-Sudarshan equation \cite{lindblad1976generators,gorini1976completely}, which stands for the dissipative dynamics of the state $\rho_t$ of the system:
\begin{align}
    \partial_t \rho_t & = \mathbb{L}_t \rho_t \, ,\label{eq12} \\ 
    \mathbb{L}_t X &= -i \left[H_{\lambda_t}, X \right] \nonumber \\&~~- \frac{1}{2}
    \sum_{\sigma = \pm}
    \left(
    \left[V^{\sigma}_t X, V^{\sigma\dag}_t\right] + 
    \left[V^{\sigma}_t, X V^{\sigma\dag}_t \right]   \right) \label{eq13} \, , 
\end{align} 
where $\mathbb{L}_t$ is a time-evolution generator including the dissipative dynamics. The operator $V^{\sigma}_t$ is the quantum jump operator given by
\begin{equation}
\begin{aligned}
    V^{\pm}_t = \sqrt{\frac{1}
    {1+e^{\mp 2 \beta \Omega_t}}}
    | e_{\mp} \rangle \langle e_{\pm} | \, . \label{eq14}
\end{aligned}
\end{equation}
Here $\Omega_t$ ($=\sqrt{\lambda_t^2+\varepsilon_t^2}/2$) is the generalized Rabi frequency. The states $| e_{+} \rangle$ and $| e_{-} \rangle$ is the excited and ground state of the corresponding energies $\Omega_t$ and $-\Omega_t$, respectively. Note that this expression satisfies the quantum detailed balance condition, which guarantees that the steady state of the generator $\mathbb{L}_t$ is the instantaneous equilibrium distribution.  Besides, the distance between two states is given by the trace distance \cite{nielsen2002quantum}:
\begin{equation}
    d(\rho_1, \rho_2) = {\rm Tr}{\left[|\rho_1 - \rho_2|\right]} \, . \label{eq14-1}
\end{equation}
Note that the entropy production rate is given by  \cite{deffner2011nonequilibrium}
\begin{equation}
    \dot{\Sigma} = -{\rm Tr}\left[(\partial_t \rho_t) \ln \rho_t \right] + {\rm Tr}\left[(\partial_t \rho_t) \ln \rho_{eq,t} \right]
    \, . \label{eq15}
\end{equation}

In general, it is difficult to theoretically solve the optimization problem (\ref{eq2}) with the above entropy production.
However, when $\varepsilon =0$, the dynamics of the system is reduced to classical probabilistic process, and then, the problem is reduced to much simpler optimization problem using the calculus of variations. 
The classical probabilistic process for this case is simply given as 
\begin{equation}
    \left(
    \begin{array}{ccc}
    \dot{p}_- \\ \dot{p}_+
    \end{array}
    \right) = 
    \left(
    \begin{array}{ccc}
    -(1-\omega_t) & \omega_t \\ 
    1-\omega_t & -\omega_t
    \end{array}
    \right)
    \left(
    \begin{array}{ccc}
    p_- \\ p_+
    \end{array}
    \right)
    \, ,  \label{eq17}
\end{equation}
where $p_-$ and $p_+$ are the probability of the eigenstate $|e_-\rangle$
and $|e_+\rangle$, respectively. The function $\omega_t$ here is defined as the instantaneous equilibrium state of $|e_-\rangle$,
\begin{equation}
    \omega_t = \frac{1}{e^{-\beta\lambda_t}+1}
    \, . \label{eq18}
\end{equation}
Plugging the relation $p_+ = 1 - p_-$ into 
the master equation (\ref{eq17}), we obtain the following single equation:
\begin{equation}
    \dot{p}_- = -p_- + \omega_t \, . \label{eq19}
\end{equation}
For the classical probabilistic process obeying $\dot{p}_i =\sum_{j} W_{ij} p_j$, the entropy production rate can be written in the form \cite{benenti2017fundamental}
\begin{equation}
\dot{\Sigma} = \frac{1}{2}  \sum_{i \neq j} 
    \left(W_{ji} p_i - W_{ij} p_j\right) \ln \frac{W_{ji} p_i}{W_{ij} p_j} \, .
\end{equation} 
Using this expression,  the entropy production rate can explicitly written as
\begin{equation}
\begin{aligned}
    \dot{\Sigma} 
    &= \dot{p}_- \ln
    \frac{(1-p_-)(p_- + \dot{p}_-)}{p_- [1-(p_- + \dot{p}_-)]}
    \, . \label{eq20}
\end{aligned}
\end{equation}


In order to minimize the entropy production, we use the framework of the calculus of variations. We now define the Lagrangian $\mathcal{L}(p_-,\dot{p}_-
) :=\dot{\Sigma}$.
The extremum is then found via the standard Euler-Lagrange method, i.e., it satisfies
\begin{equation}
    \frac{\rm d}{{\rm d}t}\left(\frac{\partial \mathcal{L}(p_-,\dot{p}_-)}{\partial \dot{p}_-}\right) - \frac{\partial \mathcal{L}(p_-,\dot{p}_-)}{\partial p_-} = 0 \, . \label{eq23}
\end{equation}
After simplification, it becomes a second-order differential equation about $t$ with two boundary conditions,
\begin{equation}
\begin{aligned}
    & 2\ddot{p}_-[1-(p_-+\dot{p}_-)](p_-+\dot{p}_-) - \\
    & \dot{p}_-(\dot{p}_-+\ddot{p}_-)[1-2(p_-+\dot{p}_-)] = 0 \, , \text{~~with} \\
    & p_-(t_{\rm i}) = \frac{1}{e^{-\beta\lambda(t_{\rm i})}+1} \text{~~and~~} 
    p_-(t_{\rm f}) = \frac{1}{e^{-\beta\lambda(t_{\rm f})}+1} \, . \label{eq24}
\end{aligned}
\end{equation}

Furthermore, we notice that if the distance between the initial state and the final state is too large, Eq. (\ref{eq24}) cannot provide a real solution of $p_-$ when the given time period $\tau$ is fixed. We calculate the upper bound of the distance here. For simplifying the calculation, we assume that the initial state $p_-(t_{\rm i})$ is fixed to be $0.5$ (i.e., $\lambda(t_{\rm i})$ = 0.0), and we only need to determine the upper bound of $p_-(t_{\rm f})$. In the following derivation, we use $p$, $p_{\rm eq,i}$ and $p_{\rm eq, f}$ to represent $p_{-}$, $p_-(t_{\rm i})$ and $p_-(t_{\rm f})$, respectively.

We start from the integration form of Eq. (\ref{eq23}):
\begin{equation}
    {\cal L} (p,\dot{p}) - \dot{p} \frac{\partial {\cal L}}{\partial \dot{p}} = -K \, , \label{eq24-1}
\end{equation}
where $K$ is a constant. After simplification, we have,
\begin{equation}
    \frac{\dot{p}^2}{(p+\dot{p})[1-(p+\dot{p})]} = K \, . \label{eq24-2}
\end{equation}
Noticing that the term $(p+\dot{p})/[1-(p + \dot{p})]$ in Eq. (\ref{eq20}) must be positive because the 
entropy production rate $\dot{\Sigma}$ is a real value, we find that $K$ is non-negative according to Eq. (\ref{eq24-2}). Thus, Eq. (\ref{eq24-2}) is a quadratic equation. The solution of it can be expressed as,
\begin{equation}
    \dot{p} = \frac{(1-2p)K \pm \sqrt{\Delta}}{2(K+1)} \, , \label{eq24-3}
\end{equation}
where $\Delta = K^2 + 4Kp(1-p)$.

In our setup, the controllable parameter $\lambda(t)$ is finite and non-negative (because it is the energy of the $| e_{+} \rangle$), which ensures that the instantaneous equilibrium state $p_{\rm eq}(\lambda(t))$ is in the range of $[0.5, 1)$. Moreover, $p_{\rm eq,i}$ is set to be $0.5$, and $p_{\rm eq,f}$ should be set to be larger than $p_{\rm eq,i}$. Hence, we have the relation $p_{\rm eq,f} > p_{\rm eq,i} = 0.5$. To satisfy this, the equation (\ref{eq24-3}) takes the "$+$" branch. The exact value of $K$ can be calculated by integrating both sides of Eq.(\ref{eq24-3}):
\begin{equation}
\begin{aligned}
    F(p_{\rm eq,f},K) - F(p_{\rm eq,i},K) = \tau \, , \label{eq24-4}
\end{aligned}
\end{equation}
where
\begin{equation}
\begin{aligned}
    F(p,K) = & \int {\rm d}p \, \frac{2(K+1)}{(1-2p)K + \sqrt{\Delta}} \\
    = & -\ln{(1-p)} + \frac{1}{2} \ln{\left[\frac{2(1-p)+K+\sqrt{\Delta}}{2p+K+\sqrt{\Delta}}\right]} \\
    & + \frac{1}{\sqrt{K}}\arctan{\left[\frac{(1-2p)\sqrt{K}}{\sqrt{\Delta}}\right]} \, . \label{eq24-5}
\end{aligned}
\end{equation}

Noticing that the first derivative of $p_{\rm eq,f}$ with respect to $K$ is non-negative (See the detail in Appendix B), we deduce that $p_{\rm eq,f}$ monotonically increases as $K$ increases. When $K$ approaches to infinity, we can calculate the upper bound of $p_{\rm eq,f}$:
\begin{equation}
\begin{aligned}
    & \lim_{K \rightarrow +\infty} F(p_{\rm eq,f},K) - \lim_{K \rightarrow +\infty} F(p_{\rm eq,i},K) = \tau \\
    \Rightarrow ~~ & \ln{(1-p_{\rm eq,i})} - \ln{(1-p_{\rm eq,f})} = \tau \\
    \Rightarrow ~~ & p_{\rm eq,f} = 1-(1-p_{\rm eq,i})e^{-\tau} \, . \label{eq24-7}
\end{aligned}
\end{equation}

Because we have already set $p_{\rm eq,i} = 0.5$ and $\tau = 1$ in our calculation. Hence, $\max{(p_{\rm eq,f})} = 1 - 1/2e \approx 0.82$. The corresponding maximum value of $\lambda(t_{\rm f})$ is
\begin{equation}
\begin{aligned}
    \max{\lambda(t_{\rm f})} = -\ln{\left[\frac{1}{p_{\rm eq,f}}- 1\right]} \approx 1.49 \, . \label{eq24-8}
\end{aligned}
\end{equation}

Choosing the suitable value of $\lambda(t_{\rm f})$ according to (\ref{eq24-8}) and solving the equation (\ref{eq24}) numerically, one can obtain the theoretical optimal path that minimizes the entropy production during the transition between two equilibrium states in the classical case. This will be used to check the performance of our RL method. Please see Appendix B for the details of the calculation. We also remark that the trace distance (\ref{eq14-1}) is reduced to $L_1$-norm distance in the classical case:
\begin{equation}
    d(\rho_1, \rho_2) = \sum_{\sigma = \pm} |p_{1,\sigma} - p_{2,\sigma}| \, . \label{eq21}
\end{equation}

For the cases that $\varepsilon$ continuously changes with time, we will show that our RL method can also find such optimal path in the subsequent section. 

\begin{figure*}[t]
\includegraphics[width=1\textwidth]{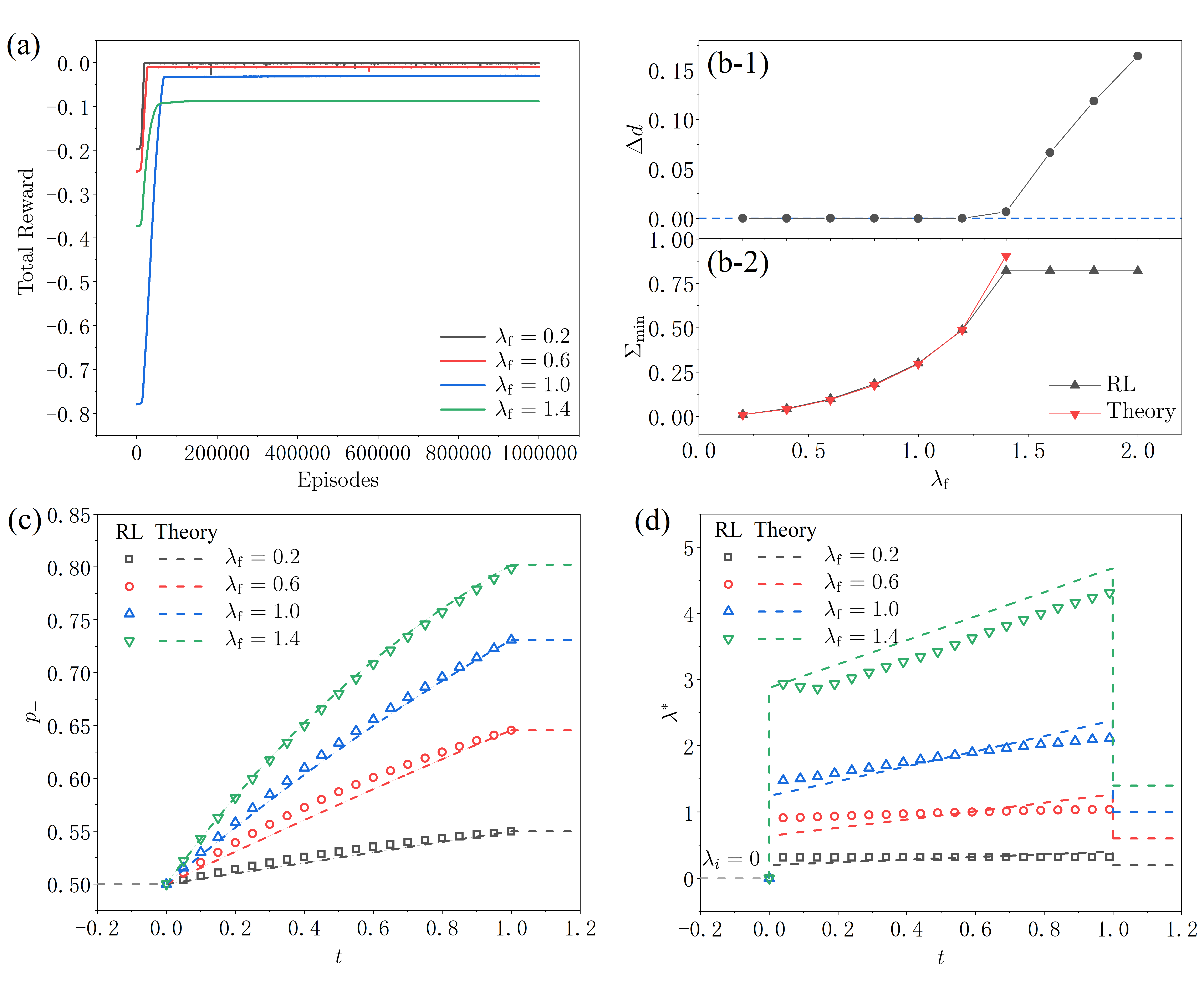}
\caption{\label{fig:1} Results for finding the optimal path that minimizes the entropy production during the transition between two equilibrium states for the classical two-level system by the reinforcement learning (RL) method. (a) shows the total rewards in RL setups increases and converges to a constant as the number of episodes increases from $0$ to $10^6$. (b) shows how the distance $\Delta d$ between the target state and the final state obtained by the RL method ((b-1)), and how the corresponding minimum entropy production $\Sigma_{\rm min}$ ((b-2)) behaves when given the different target values of $\lambda_{\rm f}$. Red down-triangles here are the theoretical results obtained by numerically solving Eq.(\ref{eq24}). (c) provides the optimal evolution of the probability of staying at the state $|e_-\rangle$ for the given different target value of $\lambda_{\rm f}$ denoted by different symbols. (d) shows the optimal protocols of the controllable parameter $\lambda(t)$ for different $\lambda_{\rm f}$s by different symbols. In (c) and (d), theoretical results are also provided for comparison denoted by dash lines with corresponding colors. }.
\end{figure*}

\begin{figure*}[t]
\includegraphics[width=1\textwidth]{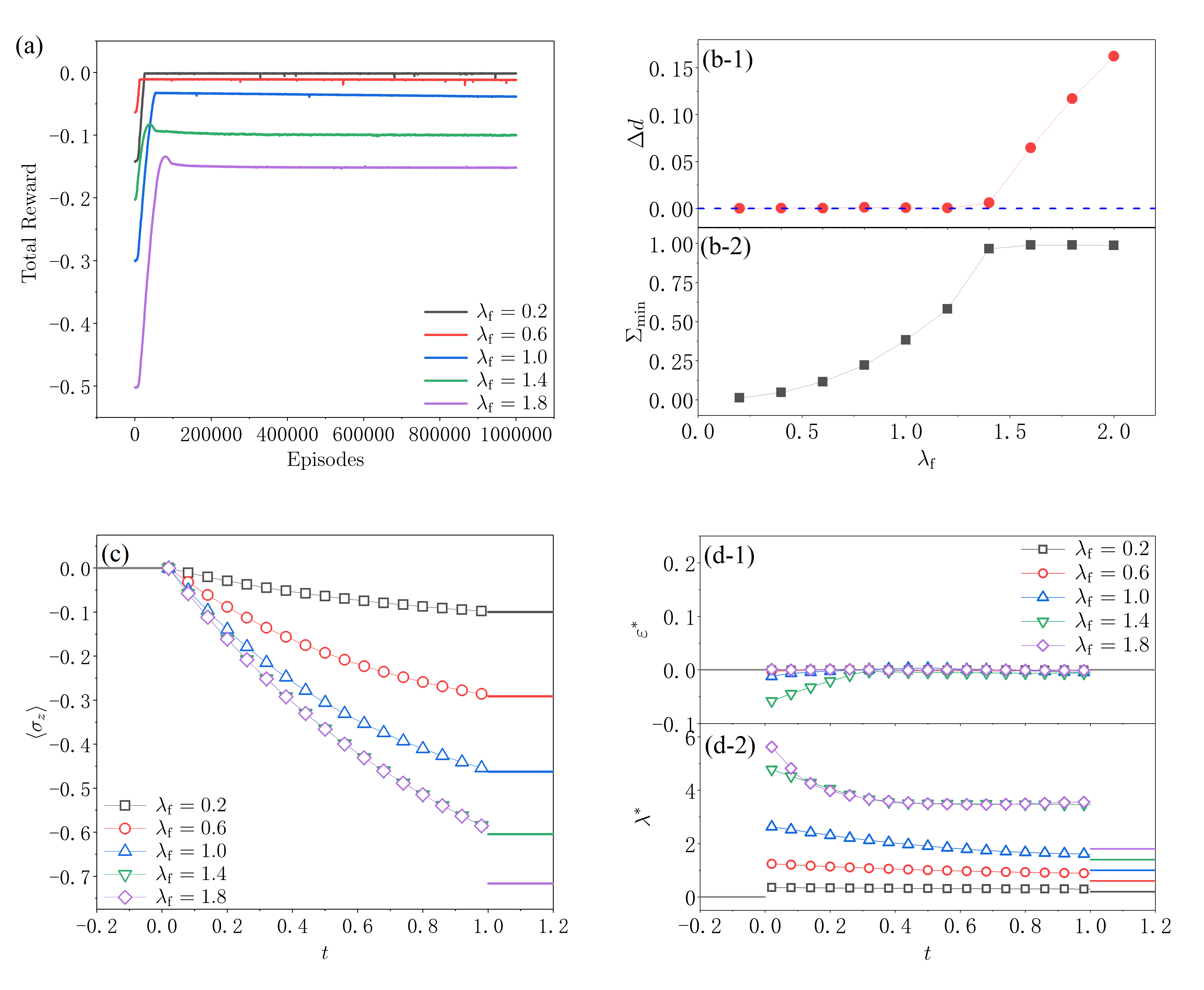}
\caption{\label{fig:2} Results for finding the optimal path that minimizes the entropy production during the transition between two equilibrium points for the quantum two-level system where $(\varepsilon (t_{\rm i}) , \varepsilon (t_{\rm f}) )=(0.0,0.0)$. (a) shows the total rewards in RL setups increases and converges to a constant as the number of episodes increases from $0$ to $10^6$. (b) shows how the distance $\Delta d$ between the target state and the final state obtained by the RL method ((b-1)), and how the corresponding minimum entropy production $\Sigma_{\rm min}$ ((b-2)) behaves when given the different target value $\lambda_{\rm f}$. Blue dash line in (b-1) denotes for the value $\Delta d = 0$. (c) provides the optimal evolution of expected value of $\sigma_z$ for the given different target value of $\lambda_{\rm f}$ denoted by different symbols. (d-1) and (d-2) respectively shows the optimal controllable parameter $\varepsilon^*$ and $\lambda^*$ for different $\lambda_f$ by different symbols. Gray lines in (c) and (d) denote the corresponding quantities at the initial state, while color lines corresponding to the color of RL results denote the corresponding quantities at the target states.}.
\end{figure*}

\begin{figure*}[t]
\includegraphics[width=1\textwidth]{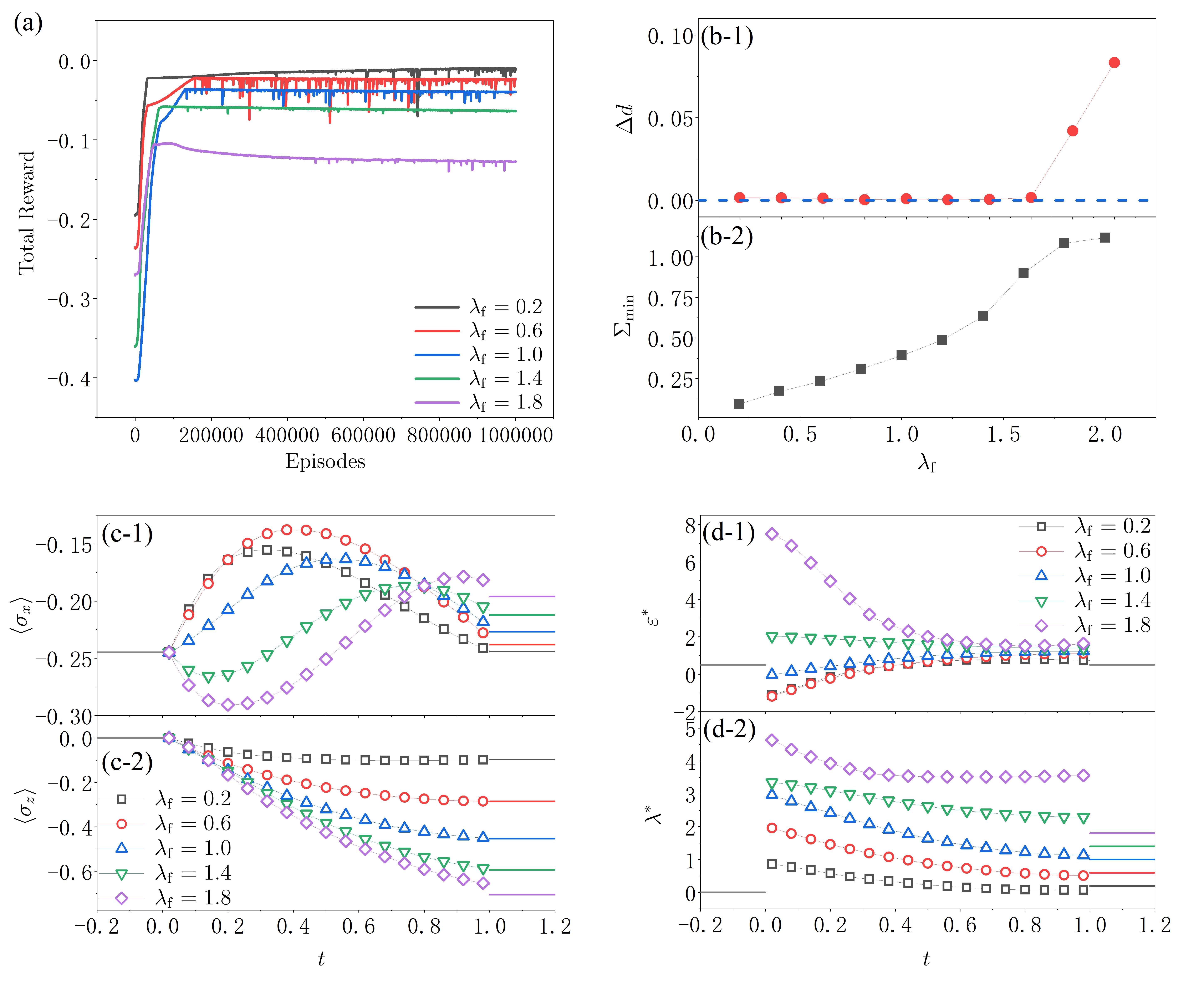}
\caption{\label{fig:3} Results for finding the optimal path that minimizes the entropy production during the transition between two equilibrium points for the quantum two-level system where 
$(\varepsilon (t_{\rm i}), \varepsilon (t_{\rm f}) )=(0.5,0.5)$. (a) shows the total rewards in RL setups increases and converges to a constant as the number of episodes increases from $0$ to $10^6$. (b) shows how the distance $\Delta d$ between the target state and the final state obtained by the RL method ((b-1)), and how the corresponding minimum entropy production $\Sigma_{\rm min}$ ((b-2)) behaves when given the different target value $\lambda_{\rm f}$. Blue dash line in (b-1) denotes for the value $\Delta d = 0$. (c-1) and (c-2) respectively provides the optimal evolution of expected value of $\sigma_x$ and $\sigma_z$ for the given different target value of $\lambda_{\rm f}$ denoted by different symbols. (d-1) and (d-2) respectively shows the optimal controllable parameter $\varepsilon^*$ and $\lambda^*$ for different $\lambda_{\rm f}$s by different symbols. Gray lines in (c) and (d) denote the corresponding quantities at the initial state, while color lines corresponding to the color of RL results denote the corresponding quantities at the target states.}.
\end{figure*}

\section{\label{sec:level5}Numerical experiment}

In this section, we present the numerical results obtained by the above RL method and we show that it works well for our problem (\ref{eq2}). Here, we consider both classical and quantum cases. 

For the classical case, the time-dependent controllable parameter is only one parameter, i.e., $\Lambda_t = \lambda(t)$ ($\varepsilon $ is always $0$), which leads to that the policy function $\pi(a|s,\bm{\theta})$ is an one-dimensional normal distribution (Eq.(\ref{eq10})). Specifically, as illustrated above, if the boundary condition is properly chosen, the theoretical result can be precisely obtained by numerically solving Eq. (\ref{eq24}). Hence, we can compare results by RL method with the exact one. Through the classical case, we can establish a benchmark for our RL method. 

For the quantum case, as two controllable parameters are considered (i.e., $\Lambda_t = (\varepsilon_t, \lambda_t)$), Eq. (\ref{eq10}) have to be a two-dimensional normal distribution. Specifically,
two different situations are considered here, i.e., the situations of 
$(\varepsilon(t_{\rm i}), \varepsilon(t_{\rm f})) = (0.0,0.0)$ and $(0.5,0.5)$. For both classical and quantum cases, the initial $\lambda_{\rm i}$ are set to be $0.0$. By changing the value of $\lambda_{\rm f}$ and fixing the time duration $\tau$ ($=t_{\rm f} - t_{\rm i}$) to be $1.0$, we show optimal paths for different final states.

The results for the classical case are shown in the Fig. \ref{fig:1} and for the quantum cases are shown in the Figs. \ref{fig:2} and \ref{fig:3}.

\subsection{Classical case}
We first discuss results for the classical case. In Fig. \ref{fig:1} (a), one can notice that the cumulative reward obtained by the agent can relatively well converge after sufficient number of episodes. When the total reward is maximized, we first check the distance between the final state obtained by our RL method and the target state which is specified by the given $\lambda_{\rm f}$. Note that the mathematical definition of $\Delta d$ is given in Eq.(\ref{eq2}). If $\Delta d$ approximately equals $0$, we regard that the corresponding target state can be reached from the initial state with the minimum entropy production. 
Note that for the parameter $\lambda_{\rm f}$ with very large distance, one may not be able to reach $\rho_{\rm eq,f}$ within a given time $\tau$. 
Indeed, as shown in Fig. \ref{fig:1} (b-1), the RL method gives an obvious turning point that $\lambda_{\rm f} \approx 1.4$ which separates the reachable and unreachable target states, which corresponds well with our theoretical analysis in (\ref{eq24-8}). Simultaneously, we notice that the corresponding minimum entropy production $\Sigma_{\rm min}$ monotonically increases with $\lambda_{\rm f}$ until it reaches around this turning point in Fig. \ref{fig:1}.
This can make sense because the dissipated energy, which is determined by the entropy production, will increase as the target state is going further and further to the initial state under the condition that the evolution time $\tau$ is fixed. In other words, there is a trade-off between the entropy production and the distance between the initial and target state: the system has to sacrifice some dissipated energy into the environment to reach the target state until that the target state is too far to be reachable. In our example, when $\lambda_{\rm f} \gtrsim 1.4$, no matter how we control $\lambda(t)$ to increase the entropy production, target states cannot be achieved. Thus, the plateau in Fig. 1 (b-2) shows that the RL method can only control the system to the state that sacrifices the maximum entropy production bounded by the evolution time $\tau$. This trade-off phenomenon reminds us of the speed limit in classical systems, which is reported in Ref. \cite{shiraishi2018speed}. 

The protocols of target states that $\lambda_{\rm f} \lesssim 1.4$ are exactly the optimal ScI protocols that we intended to find in the problem (\ref{eq2}). In Fig. \ref{fig:1} (c), the evolution of the probability of staying at the state $|e_-\rangle$ for each $\lambda_{\rm f}$ is performed. One can see that under the control of our RL method, $p_-$ continuously and precisely varies with time from the fixed initial state to the designed target state.

The corresponding optimal protocols ($\lambda^*(t)$) are shown in Fig. \ref{fig:1} (d). A priori, one might expect that $\lambda^*(t)$ connecting the given initial and ﬁnal values are smoothly. However, we notice that jumps appears at the initial and final time of $\lambda^*(t)$. These jumps are also noticed when finding the optimal transition between two equilibrium states with minimum work in both classical Brownian particle model and the single-qubit model \cite{band1982finite,schmiedl2007efficiency,esposito2010finite,esposito2010quantum}. It is because of the specific mathematical structure of the equation (\ref{eq24}).

The accuracy of our RL method can be checked by the solution of Eq. (\ref{eq24}). As shown in Fig. \ref{fig:1} (b-2), (c) and (d), one can notice that the RL can relatively matches the theoretical results, which means our RL method is reliable. 
Besides, we cannot obtain the theoretical results for $\lambda_{\rm f} \gtrsim 1.4$ because Eq. (\ref{eq24}) has no real solution. Intriguingly, the RL method correctly produce the unreachable parameter regime.  

\subsection{Quantum case}
Nice performances of our RL method in the classical case encourage us to study the quantum case. Considering that the state of the quantum two-level system is a four-dimensional vector (vector representation of density matrix $\rho$) no longer a two-dimensional one, we think that an additional controllable parameter is needed because $\lambda(t)$ in Eq. (\ref{eq11}) can only control the diagonal terms of the Hamiltonian. Thus, the controllable vector is two-dimensional in the quantum case, i.e., $\Lambda_t = (\varepsilon_t, \lambda_t)$.

For the situation that $(\varepsilon_{\rm i}, \varepsilon_{\rm f}) = (0.0, 0.0)$, the RL method gives similar results with the classical case shown in Fig. \ref{fig:2}. Comparing the value of $\Sigma_{\rm min}$ in Fig. \ref{fig:1} (b-2) and Fig. \ref{fig:2} (b-2), we find that extra energy is sacrificed to overcome the quantum effect caused by $\varepsilon(t)$. To avoid this extra energy cost, the system tends to stay around the quasi-classical limit, which is the reason why the optimal $\varepsilon^*$ vibrates around $0.0$ shown in Fig. \ref{fig:2} (d-1). The optimal ScI protocols $\lambda^*$ in Fig. \ref{fig:2} (d-2) is thus different from those Fig. \ref{fig:1} (d) because $\varepsilon^*$ is not constantly to be zero. Based on this, we deduce that the optimal protocol in this situation may not be unique. Besides, the performance of the optimal paths found by the RL method is shown in Fig. \ref{fig:2} (c). Here we shows the evolution of expected value of $\sigma_z$ for different $\lambda_{\rm f}$. Similar to the result of the classical case (Fig. 1 (c)), $\langle \sigma_z \rangle$ continuously varies with time and can reach the corresponding reachable target state. Our RL method also works well for this situation.

Similar phenomena also appears in the pure quantum situation where we set $(\varepsilon_{\rm i}, \varepsilon_{\rm f}) = (0.5, 0.5)$. In this situation, as shown in Fig. \ref{fig:3} (b-1), the turning point moves to around $1.6$ because the fixed initial state is changed when $\varepsilon_{\rm i}$ becomes $0.5$. The target state for each $\lambda_{\rm f}$ is no longer the same with that in the classical case, which should be calculated by Eq. (\ref{eq1}). Again, compared to (b-2) of Fig. \ref{fig:2} and Fig. \ref{fig:3}, the results here show more significant behavior to overcome the quantum effect by sacrificing more entropy production than
cases above. The expected values of $\sigma_x$ and $\sigma_z$ are respectively applied to check the performance of our RL method, and they also work well as shown in Fig. \ref{fig:3} (c-1) and (c-2). 

\section{\label{sec:level6} Conclusion}
In summary, we propose a RL method based on techniques of policy gradient to find the optimal protocol for the controllable parameter that minimize the total entropy production during the finite-time transition between two equilibrium states. Both classical and quantum two-level systems are considered. The optimal protocols for each system are successfully obtained by our method. Especially, for the classical case, by comparing RL results with theoretical results, we show our RL method is reliable for this optimization problem. When searching for these optimal protocols, we find that: i) Jumps at initial and final time on the protocol of controllable systems appear because of the special type of variational differential equations. While it has been clarified in classical cases, it is still worthwhile to clarify the jumps in quantum case theoretically. ii) accelerating transitions between two equilibrium states requires a part of extra dissipated energy which increases the entropy production. For a given finite time period, there is a upper bound on the entropy production beyond which the target state cannot be reached. This phenomena is related to the speed limit in classical and quantum systems. While the speed limit is not a tight bound, a tight bound is further investigated. iii) Our results shows that compared to the classical two-level system, the quantum system need to sacrifice more energy to overcome the quantum effect and reach the target state.

Besides, the advantage of our RL method is that it can be generalized to other optimal-protocol-seeking problems, especially shortcuts between equilibrium states with maximizing or minimizing some other physical quantities. We believe our method may also be applied to find the optimal protocols of transitions between two nonequilibrium steady states.

\begin{acknowledgments}
We thank K. Saito for his useful discussions and encouragement. This work is supported by RIKEN Junior Research Associate.
\end{acknowledgments}

\begin{widetext}
\appendix

\section{Details of the reinforcement learning method}
We provide here some technical details and hyperparameters used in this work. 

Our neural network has 3 hidden layers and each layer contains 100 neurons with Rectiﬁed Linear Unit activation function. We set the activation function of the ﬁnal layer to be a hyperbolic tangent times a constant so that we can control the maximum value of the output by changing the value of this constant. Here we set it to be $10$. The training is performed using the Adam optimizer \cite{kingma2014adam}, and the learning rate $\alpha$ is set to be $10^{-5}$ for the classical case, while $2\times10^{-6}$ for the quantum case. The covariance matrix $\Sigma$ in Eq. (\ref{eq10}) is a scalar and we set it to be $0.01$ for the classical case. And it is a two-dimensional matrix,
\begin{equation}
\Sigma_a = \left(
\begin{matrix}
    0.01 & 0 \\
    0 & 0.01 
\end{matrix}
\right)
\end{equation}
for the quantum case.
We also subtracted a baseline to the expected return for each step of the RL episodes, i.e., $G_j \gets G_j-w$, where $w$ is initially a random number and can be renewed by $\Delta w = \alpha_w G_j$ in each step of the RL episodes with the learning rate $\alpha_w = 10^{-4}$. This method usually leads to better convergence properties \cite{sutton2018reinforcement}.
All calculations of nerual networks are based on Pytorch tools \cite{paszke2019pytorch}. The evolution of quantum states and the numerical calculation of Lindablad equation are calculated by QuTip toolbox in Python \cite{johansson2012qutip}.

\section{Calculation of the upper bound of the final state in the classical two-level system}
We provide the details of the calculation of the upper bound of final state $p_{\rm eq,f}$ in the classical two-level system. 

As illustrated in the main text, we start from the integration form of Euler-Lagrange equation (\ref{eq24-1}):
\begin{equation}
    {\cal L} (p,\dot{p}) - \dot{p} \frac{\partial {\cal L}}{\partial \dot{p}} = -K \, . \label{eqa2-1}
\end{equation}
Here, according to the definition of entropy production (\ref{eq20}), we have
\begin{equation}
\begin{aligned}
\frac{\partial {\cal L}}{\partial \dot{p}} = & \frac{\partial }{\partial \dot{p}}\left\{ \dot{p} \ln{\frac{(1-p)(p+\dot{p})}{p[1-(p+\dot{p})]}} \right\} \\
= & \ln{\frac{(1-p)(p+\dot{p})}{p[1-(p+\dot{p})]}} \\ & + \dot{p} \cdot \frac{p[1-(p+\dot{p})]}{(1-p)(p+\dot{p})} \cdot \frac{(1-p)p[1-(p+\dot{p})] - (1-p)(p+\dot{p})(-p)}{\{p[1-(p+\dot{p})]\}^2} \\
= & \ln{\frac{(1-p)(p+\dot{p})}{p[1-(p+\dot{p})]}} + \dot{p} \cdot \frac{1}{(1-p)(p+\dot{p})} \cdot \frac{(1-p)p}{p[1-(p+\dot{p})]} \\
= & \ln{\frac{(1-p)(p+\dot{p})}{p[1-(p+\dot{p})]}} + \frac{\dot{p}}{(p+\dot{p})[1-(p+\dot{p})]} \, . \label{eqa2-2}
\end{aligned}
\end{equation}
Plugging (\ref{eqa2-2}) into (\ref{eqa2-1}), we obtain
\begin{equation}
    \frac{\dot{p}^2}{(p+\dot{p})[1-(p+\dot{p})]} = K \, . \label{eqa2-3}
\end{equation}
Because the entropy production rate $\dot{\Sigma}$ is a real number, the term $\frac{(p+\dot{p})}{1-(p+\dot{p})}$ must be a positive value. Hence, $K \geq 0$.

After simplification, we have
\begin{equation}
    (K+1)\dot{p}^2 + K(2p-1)\dot{p} - Kp(1-p) = 0 \, . \label{eqa2-4}
\end{equation}
The solution can be expressed as,
\begin{equation}
    \dot{p} = \frac{(1-2p)K \pm \sqrt{\Delta}}{2(K+1)} \, , \label{eqa2-5}
\end{equation}
where $\Delta = K^2 + 4Kp(1-p)$.

Now we explain the physical meaning of $K$. By inserting the master equation (\ref{eq19}) into Eq. (\ref{eqa2-3}), we obtain
\begin{equation}
\begin{aligned}
     & \frac{(p-\omega_t)^2}{\omega_t (1-\omega_t)} = K \\
     \Rightarrow ~~ & p = \omega_t (1 \pm \sqrt{\frac{1-\omega_t}{\omega_t}} \sqrt{K}) \, . \label{eqa2-6}
\end{aligned}
\end{equation}
Because $\omega_t$ is the instantaneous equilibrium state, we notice that the system approaches to the quasistatic limit as $K$ approaches to $0$. We think that $K$ is a quantity that measures how far the state of the system varies from the quasistatic limit. In our setup, we limit $\lambda(t) \geq 0$. Hence, we have the relation $p_{\rm eq,f} > p_{\rm eq,i} \geq 0.5$. Thus, the equation (\ref{eqa2-5}) and (\ref{eqa2-6}) takes the "$+$" branch. 
The exact value of $K$ can be calculated by integrating both sides of Eq.(\ref{eq10}):
\begin{equation}
\begin{aligned}
    & \int_{p_{\rm eq,i}}^{p_{\rm eq,f}} {\rm d}p \, \frac{2(K+1)}{(1-2p)K + \sqrt{\Delta}} = \tau \\
    \Rightarrow ~~ & F(p_{\rm eq,f},K) - F(p_{\rm eq,i},K) = \tau \, ,
\end{aligned}
\end{equation}
where
\begin{equation}
\begin{aligned}
    F(p,K) = & \int {\rm d}p \, \frac{2(K+1)}{(1-2p)K + \sqrt{\Delta}} \\
    = & -\ln{(1-p)} + \frac{1}{2} \ln{\left[\frac{2(1-p)+K+\sqrt{\Delta}}{2p+K+\sqrt{\Delta}}\right]} + \frac{1}{\sqrt{K}}\arctan{\left[\frac{(1-2p)\sqrt{K}}{\sqrt{\Delta}}\right]} \, .
\end{aligned}
\end{equation}
The value of $K$ is determined once $p_{\rm eq,i}$, $p_{\rm eq,f}$ and $\tau$ are fixed.

To find the maximum $p_{\rm eq,f}$ when given $p_{\rm eq,i}$ and $\tau$, we first calculate the derivatives of $K$ and $p_{\rm eq,f}$.

We define $G(p_{\rm eq,f},K|p_{\rm eq,i},\tau) = F(p_{\rm eq,f},K) - F(p_{\rm eq,i},K) - \tau$. The differential of $G$ is calculated by
\begin{equation}
\begin{aligned}
    {\rm d}G = & \frac{\partial G}{\partial p_{\rm eq,f}} {\rm d}p_{\rm eq,f} + \frac{\partial G}{\partial K} {\rm d}K 
\end{aligned}
\end{equation}
Here, we have
\begin{equation}
\begin{aligned}
    \frac{\partial G}{\partial p_{\rm eq,f}} = \frac{\partial F}{\partial p_{\rm eq,f}} = \frac{2(K+1)}{(1-2p_{\rm eq,f})K + \sqrt{\Delta_{\rm f}}} \, ,
\end{aligned}
\end{equation}
and
\begin{equation}
\begin{aligned}
    \frac{\partial G}{\partial K} = &\frac{\partial }{\partial K} \int_{p_{\rm eq,i}}^{p_{\rm eq,f}} {\rm d}p \, F(p,K)  \\
    = & \int_{p_{\rm eq,i}}^{p_{\rm eq,f}} {\rm d}p \, \frac{\partial F}{\partial K} \\
    = & \int_{p_{\rm eq,i}}^{p_{\rm eq,f}} {\rm d}p \, \frac{\partial }{\partial K} \left[\frac{2(K+1)}{(1-2p)K+\sqrt{\Delta}}\right] \\
    = & - \int_{p_{\rm eq,i}}^{p_{\rm eq,f}} {\rm d}p \, \frac{1}{K\sqrt{\Delta}} \\
    = &  \frac{1}{{2K\sqrt{K}}} \left\{ \arctan{\left[\frac{(2p_{\rm eq,i}-1)\sqrt{K}}{\sqrt{\Delta_{\rm i}}}\right]} - \arctan{\left[\frac{(2p_{\rm eq,f}-1)\sqrt{K}}{\sqrt{\Delta_{\rm f}}}\right]}\right\} \, .
\end{aligned}
\end{equation}

Hence, the derivative of $p_{\rm eq,f}$ with respect to $K$ is then obtained,
\begin{equation}
\begin{aligned}
    \frac{{\rm d} p_{\rm eq,f}}{{\rm d} K} = & -\frac{\frac{\partial G}{\partial K}}{\frac{\partial G}{\partial p_{\rm eq,f}}} \\
    = & \frac{(1-2p_{\rm eq,f})K + \sqrt{\Delta_{\rm f}}}{4K\sqrt{K}(K+1)} \\
    & \times \left\{ \arctan{\left[\frac{(2p_{\rm eq,f}-1)\sqrt{K}}{\sqrt{\Delta_{\rm f}}}\right]} - \arctan{\left[\frac{(2p_{\rm eq,i}-1)\sqrt{K}}{\sqrt{\Delta_{\rm i}}}\right]} \right\} \, .
\end{aligned}
\end{equation}
Because
\begin{equation}
\begin{aligned}
    (1-2p_{\rm eq,f})K + \sqrt{\Delta_{\rm f}} = & \frac{(\sqrt{\Delta_{\rm f}} - (2p_{\rm eq,f}-1)K)(\sqrt{\Delta_{\rm f}} + (2p_{\rm eq,f}-1)K)}{\sqrt{\Delta_{\rm f}} + (2p_{\rm eq,f}-1)K} \\
    = & \frac{\Delta_{\rm f} - (2p_{\rm eq,f}-1)^2 K^2}{\sqrt{\Delta_{\rm f}} + (2p_{\rm eq,f}-1)K} \\
    = & \frac{4 p_{\rm eq,f} (1-p_{\rm eq,f})K(K+1)}{\sqrt{\Delta_{\rm f}} + (2p_{\rm eq,f}-1)K} \geq 0 \, ,
\end{aligned}
\end{equation}
and
\begin{equation}
\begin{aligned}
    & \frac{(2p_{\rm eq,f}-1)\sqrt{K}}{\sqrt{\Delta_{\rm f}}} - \frac{(2p_{\rm eq,i}-1)\sqrt{K}}{\sqrt{\Delta_{\rm i}}} \\
    = &
    \sqrt{\frac{k}{\Delta_{\rm i}\Delta_{\rm f}}} \left[(2p_{\rm eq,f}-1)\Delta_{\rm i} - (2p_{\rm eq,i}-1)\Delta_{\rm f}\right] \\
    = & \sqrt{\frac{k}{\Delta_{\rm i}\Delta_{\rm f}}} \frac{[(2p_{\rm eq,f}-1)\Delta_{\rm i} - (2p_{\rm eq,i}-1)\Delta_{\rm f}][(2p_{\rm eq,f}-1)\Delta_{\rm i} + (2p_{\rm eq,i}-1)\Delta_{\rm f}]} {[(2p_{\rm eq,f}-1)\Delta_{\rm i} + (2p_{\rm eq,i}-1)\Delta_{\rm f}]} \\
    = & \sqrt{\frac{k}{\Delta_{\rm i}\Delta_{\rm f}}} \frac{[4p_{\rm eq,i}(1-p_{\rm eq,i})-4p_{\rm eq,f}(1-p_{\rm eq,f})] (K^2 + 1)}{[(2p_{\rm eq,f}-1)\Delta_{\rm i} + (2p_{\rm eq,i}-1)\Delta_{\rm f}]} \geq 0 \, ,
\end{aligned}
\end{equation}
the derivative of $p_{\rm eq,f}$ with respect to $K$ is non-negative. Thus, $p_{\rm eq,f}$ monotonically increases as $K$ increases. When $K$ approaches to $\infty$, we have the upper bound of $p_{\rm eq,f}$
\begin{equation}
\begin{aligned}
    & \lim_{K \rightarrow \infty} F(p_{\rm eq,f},K) - \lim_{K \rightarrow \infty} F(p_{\rm eq,i},K) = \tau \\
    \Rightarrow ~~ & \ln{(1-p_{\rm eq,i})} - \ln{(1-p_{\rm eq,f})} = \tau \\
    \Rightarrow ~~ & p_{\rm eq,f} = 1-(1-p_{\rm eq,i})e^{-\tau}
\end{aligned}
\end{equation}

In our setup, we set $p_{\rm eq,i} = 0.5$ and $\tau = 1$. Hence, $\max{(p_{\rm eq,f})} = 1 - 1/2e \approx 0.82$. The corresponding turning point of $\lambda_{\rm f}$ is
\begin{equation}
\begin{aligned}
    \lambda_{\rm f} = -\ln{\left[\frac{1}{p_{\rm eq,f}}- 1\right]} \approx 1.49 \, .
\end{aligned}
\end{equation}
The results given by our RL method is very close to the theoretical one, where $max(p_{\rm eq, f}) \approx 0.80$ and $\lambda_{\rm f} \approx 1.38$.
\end{widetext}

\bibliography{references}

\end{document}